\documentclass[twocolumn,showpacs,preprintnumbers,amsmath,amssymb]{revtex4}
\usepackage{graphicx}
\usepackage{dcolumn} 
\usepackage{bm}      

\begin{document}
\title{Transverse Mode Coupling Instability with Space Charge}
\author{V. Balbekov}
\email{balbekov@fnal.gov}
\affiliation {Fermi National Accelerator Laboratory\\
 P.O. Box 500, Batavia, Illinois 60510, email balbekov@fnal.gov} 
\date{\today}

\begin{abstract}

Transverse mode coupling instability of a bunch with space charge and wake field
is considered within the frameworks of the boxcar model.
Eigenfunctions of the bunch without wake are used as a basis for solution 
of the equations with the wake field included.
Dispersion equation for the bunch eigentunes is presented in the form of 
an infinite continued fraction and also as the recursive relation with 
arbitrary number of the basis functions.
It is shown that the influence of space charge on the instability essentially
depends on the wake sign.
In particular, threshold of the negative wake increases in absolute value 
until the space charge tune shift is rather small, and goes down at the
higher space charge.
The explanation is developed by analysis of the bunch spectrum.
	
\end{abstract}

\maketitle
%

\section{INTRODUCTION}

%
Transverse Mode Coupling Instability has been observed first in PETRA and 
has been explained by Kohaupt on the base of two-particle model \cite{KO}. 
Many papers concerning this effect were published later, 
including handbooks and surveys (see e.g. \cite{NG}).
It is established that the instability occurs by a coalescence
of two neighboring head-tail modes due to the bunch wake field.  

TMCI with space charge was considered first by Blaskiewicz \cite{BL1}.
The main conclusion which has been done in the paper and many times quoted 
is that SC enhances the TMCI threshold that is improve the beam stability.

The problem has been investigated later in the case of high space charge
when the tune shift significantly exceeds synchrotron tune \cite{BUR, BA1}.
It was confirmed that SC can stabilize TMCI suppressing the instability 
in considered limiting case (``vanished TMCI'' \cite{BUR}). 
However, it was shown in Ref.~\cite{BA1} that this statement can be true only 
for negative wakes whereas positive ones impair the stability resulting in a 
monotonous falling of the TMCI threshold when the SC tune shift increases.
Besides, it was suggested in Ref.~\cite{BUR} that the threshold growth of negative 
wake can also cease and turn back if the space charge tune shift exceeds synchrotron 
tune on one or two orders of value.   

Numerical simulations of the instability with a modest tune shift have confirmed these 
results on the whole demonstrating the predicted effects with positive and negative 
wakes \cite{BL2}

So-termed three-mode model has been developed in Ref.~\cite{BA2} for analytical 
description of the TMCI with chromaticity and arbitrary wake at moderate SC.
The results of this simple model are in a good agreement with those mentioned above.

However, it would be premature to declare the problem settled
because there are some lacunae and contradictions in the cited papers.

A non-monotonic dependence of the TMCI threshold and rate on the SC tune shift has been 
actually obtained in Ref.~\cite{BL1}. 
It has been shown that the stability and instability regions can change each other 
when the tune shift increases.
Besides, it appeared that the applied expansion technique does not always converge 
as more basis vectors are added.

The numerical solutions in Ref.~\cite{BL2} sometimes demonstrate a non-monotonic 
dependence of the threshold on the tune shift as well. 
However, the number of examples looks insufficient for any 
reliable conclusions about the crucial value of the shift.

The statements of papers \cite{BUR} and \cite{BA1} relating to the negative wakes 
are questionable because negative multipoles $(m<0)$ have been excluded 
from the consideration due to the basis assumptions.
Meanwhile, it is known that the coupling of the multipoles $m=0$ and $m=-1$ is 
the main factor resulting in the TMCI without space charge.

Only the case of modest space charge $\,\Delta Q_{sc}/Q_s<3\,$ has been investigated 
in Ref.~\cite{BA2} demonstrating that the TMCI threshold of negative wake goes up 
when the tune shift increases.
However, used equations allow to suggest that a sudden kink of the threshold 
curve is possible at the higher shift.
Therefore, field of application of the three-mode model is still an open question.
A verification and an explanation of these results are desirable.     

These problems are considered in the presented paper based on the analysis 
of a square bunch (the boxcar model).
An advantage of such a model is that set of its eigenmodes without wake is known 
exactly, providing very convenient basis for the consideration of the TMCI problem 
in depth. 

%

\section{BASIC EQUATIONS AND ASSUMPTIONS}

%

The terms, basic symbols and equations of Ref.~\cite{BA1} and \cite{BA2}
are used in this paper.
In particular, linear synchrotron oscillations are considered here
being characterized by amplitude $\,A\,$ and phase $\,\phi$, 
or by corresponding Cartesian coordinates: 
$$
 \theta= A\cos\phi, \qquad u=A\sin\phi.
$$
Thus $\,\theta\,$ is the azimuthal deviation of a particle from the bunch center
in the rest frame, and variable $u$ is proportional to the momentum 
deviation about the bunch central momentum 
(the proportionality coefficient plays no part in the paper).
Transverse coherent displacement of the particles in some point of the 
longitudinal phase space will be presented as real part of the function 
%
\begin{equation}
 X(A,\phi,t)=Y(A,\phi)\exp\big[-i(Q_0+\zeta)\,\theta-i\,(Q_0+\nu)\,\Omega_0t\,\big]
\end{equation}                         
%
where $\,\Omega_0\,$ is the revolution frequency, $\,Q_0$ is the central 
betatron tune,  and $\,\nu\,$ is the tune addition produced by space charge and 
wake field. 
Generally, $\,\zeta\,$~is the normalized chromaticity, however, only the case 
$\zeta=0$ will be investigated in this paper. 
Besides, we restrict the consideration by the simplest case of the wake field
which has constant value within the bunch being zero behind it.
Then the function $Y$ satisfies the equation \cite{BA1},\cite{BA2}:
%
\begin{eqnarray}
 \nu Y+i\,Q_s\frac{\partial Y}{\partial\phi}
+\frac{\Delta Q(0)\rho(\theta)}{\rho(0)}\,\big[Y(\theta,u)-\bar Y(\theta)\big]
 \nonumber \\
=2q\int_\theta^\infty\bar Y(\theta')\rho(\theta')d\theta' 
\end{eqnarray}
%
where $\,F(\theta,u)\,$ and $\,\rho(\theta)\,$ are normalized distribution function 
and corresponding linear density of the bunch, $\,Q_s\,$ is the synchrotron tune,
$\,\Delta Q(\theta)\,$ is the space charge tune shift, and $\,\bar Y(\theta)\,$ 
is the bunch displacement in usual space which can be found by the formula
%
\begin{eqnarray}
 \rho(\theta)\bar Y(\theta) = \int_{-\infty}^\infty F(\theta,u)Y(\theta,u)\,du  
\end{eqnarray}
%
Parameter $q$ is the reduced wake strength 
which is connected with the usual wake field function $W_1$ by the relation:  
%
\begin{equation}
 q =\frac{r_0R N W_1}{8\pi\beta\gamma Q_0}                     
\end{equation}
%
with $\,r_0=e^2/mc^2\,$ as the classic radius of the particle, $R$ as the 
accelerator radius, $N$ as the bunch population, $\,\beta\,$ and $\,\gamma\,$ 
as normalized velocity and energy.
More often than not, this parameter is negative, at least within the bunch. 
The typical and very known example is the resistive wall wake \cite{NG}. 
However, positive wake function is possible as well.
For example, it can be created by heavy positive ions 
arising at ionization of residual gas by proton beam.
This kind of instability has been observed in CPS \cite{HER} and U-70 \cite{GER}.
Therefore, constant wakes of both signs will be examined in this paper.

Solution of Eq.~(2) can be found by its expansion in terms of the 
eigenfunctions of corresponding uniform equation which is
%
\begin{eqnarray}
 \nu_jY_j+i\,Q_s\frac{\partial Y_j}{\partial\phi}+\frac{\Delta Q(0)\rho(\theta)}
 {\rho(0)}\,\big[Y_j(\theta,u)-\bar Y_j(\theta)\big]=0
\end{eqnarray}
%
It is easy to check that they form the orthogonal basis with the weighting 
function $\,F(\theta,u)\,$. 
Besides, we will impose the normalization condition:
%
\begin{eqnarray}
 \int\int F(\theta,u)Y^*_j(\theta,u)Y_k(\theta,u)\,d\theta du=\delta_{jk}
\end{eqnarray}
%
where the star means complex conjugation.
Then, looking for the solution of Eq.~(2) in the form
%
\begin{equation}
 Y = \sum_j C_jY_j,
\end{equation}
%
one can get the relation for the unknown coefficients $\,C_j$:               
%
\begin{eqnarray}
 \sum_j(\nu-\nu_j)C_jY_j 
=2q\sum_j C_j\int_{\theta}^\infty \bar Y_j(\theta')\rho(\theta') d\theta'
\end{eqnarray}
%
where $\,\bar Y_j\,$ and $\,Y_j\,$ are also connected by Eq.~(3). 
Multiplying Eq.~(8) by factor $\,F(\theta,u)Y^*_J(\theta,u)\,$,
integrating over $\,d\theta du\,$, and using normalization condition (6)
one can get series of equations for the coefficients $\,C$:   
%
\begin{eqnarray}
 (\nu-\nu_J)C_J  \nonumber \\
=2q\sum_j C_j\int_{-\infty}^\infty \rho(\theta)\bar Y_J^*(\theta)
 \,d\theta  \int_\theta^\infty\rho(\theta')\bar Y_j(\theta')\, d\theta',
\end{eqnarray}
%
%

\section{BOXCAR MODEL}

%

The boxcar model is characterized by following expressions for the bunch 
distribution function and its linear density:
%
\begin{subequations}
\begin{eqnarray}
 F = \frac{1}{2\pi\sqrt{1-A^2}}=\frac{1}{2\pi\sqrt{1-\theta^2-u^2}},
\end{eqnarray}
\begin{eqnarray}
 \rho(\theta)=\frac{1}{2}\qquad{\rm at}\qquad|\theta|<1
\end{eqnarray}
\end{subequations}
%
Because the eigenfunctions depend on two variables $(\theta$-$u)$ (or $A$-$\phi)$,
it is more convenient to represent $\,j\,$ as a pair of the indexes: 
$$  
 j \equiv \{n,m\},\qquad Y_j\equiv Y_{n,m}
$$
Analytical solutions of Eq.~(5) for the boxcar model have been obtained in such a form 
by Sacherer~\cite{SAH}.
The most important point is that the averaged eigenfunctions $\,\bar Y_{n,m}\,$ 
do not depend on second index being proportional to the Legendre polynomials: 
$\bar Y_{n,m}(\theta)=\bar Y_n(\theta)\propto P_n(\theta),\;n=0,\,1,\,2,\dots$.
At any $\,n,$, there are $\,n+1\,$ different eigenmodes $\,Y_{n,m}(\theta,u)\,$
which satisfy the equation
%
\begin{eqnarray}
 (\nu_{n,m}+\Delta Q)\,Y_{n,m}+i\,Q_s\frac{\partial Y_{n,m}}{\partial\phi}
=\Delta Q\,S_{n,m}P_n(\theta) 
\end{eqnarray}
%
where $\,m=n,\,n-2,\,\dots,\,-n$. 
Coefficients $\,S_{n,m}\,$ are added to the right-hand part of this equation
so that the eigenfunctions met normalization condition~(6).
Details of the calculation are represented in the Appendix, and several 
eigentunes $\,\nu_{n,m}\,$ are plotted in Fig.~1.
%
 \begin{figure}[t!]
 \hspace{-8mm}
 \includegraphics[width=75mm]{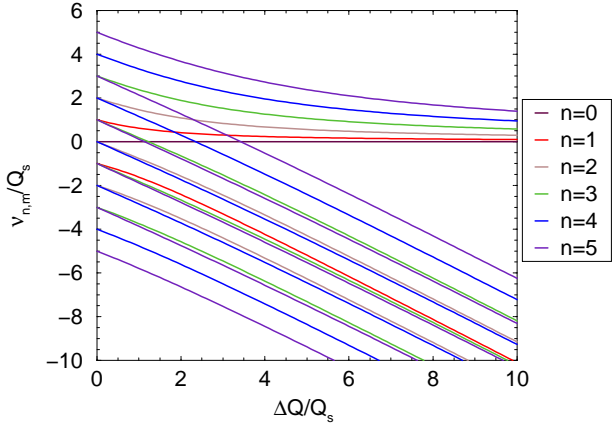}
 \caption{Eigentunes of the boxcar bunch without wake. 
 With any $\,n$, there are $\,n+1\,$ 
 eigentunes starting from different points at $\,\Delta Q=0$.}
 \end{figure}
%
It is seen that all the lines take start from the points $\,\nu_{n,m}=mQ_s\,$
at $\,\Delta Q=0\,$.
It is the commonly accepted rule to use the term 'multipole' for the collective 
synchrotron oscillations of such frequency, 
so $\,m\,$ should be treated here as the multipole number. 
Another index $\,n\,$ characterizes the eigenfunction power 
which property is usually associated with the radial mode number
when the lower number means the lower power. 
Because $\,n\ge |m|$, the mode $\,\{m,m\}$ should be treated as the lowest radial 
mode of $\,m$-th multipole.  
At $\Delta Q\ne 0$, the multipoles mix together and the eigentunes break down
into 2 groups. 
In the first of them, all tunes have positive value which tends to 0 at 
$\Delta Q\rightarrow\infty$.
It is seen that they are the lowest radial modes by the origin. 
In second group, the tunes are about 
$\nu_{n,m}\simeq mQ_s-\Delta Q\,$ being weakly dependent on the radial index $\,n$.  

With $\,\bar Y_{n,m}=S_{n,m}P_n(\theta),$ series (9) for the boxcar bunch 
obtains the form 
%
\begin{equation}
 (\nu-\nu_{NM})C_{NM}=q\,S_{NM}^*\sum_{n=0}^\infty R_{Nn}\sum_m S_{nm}C_{nm}
\end{equation}
%
with the matrix
%
\begin{equation}
 R_{Nn}=\frac{1}{2}
 \int_{-1}^1 P_N(\theta)\,d\theta\int_\theta^1 P_n(\theta')\,d\theta'
\end{equation}
%
Its small fragment is shown in Tab.~I, whereas the general form 
at $\,N\ne\,0$ is 
%
\begin{equation}
 R_{N,n}=-\frac{\delta_{N-1,n}}{(2N-1)(2N+1)}+\frac{\delta_{N+1,n}}{(2N+1)(2N+3)}
\end{equation}
%
\begin{table}[t!]
\begin{center}
\caption{Fragment of the matrix $R_{N,n}$. 
Its general form is given by Eq.~(13) and (16).}
\vspace{5mm}
\begin{tabular}{|c|c|c|c|c|c|c|}
\hline 
$n\rightarrow$ &    0    &    1    &    2     &   3    &    4   &   5    \\
\hline
$  ~ N=0 ~   $ &    1    &  ~1/3~  &     0    &   0    &    0   &~~~0~~~~\\
$  ~ N=1 ~   $ &~$-1/3$~ &    0    &    1/15  &   0    &    0   &~~~0~~~~\\
$  ~ N=2 ~   $ &    0    &$-1/15$~~&     0    &~~1/35~~&    0   &~~~0~~~~\\
$  ~ N=3 ~   $ &    0    &    0    &~$-1/35$~~&   0    &~~1/63~~&~~~0~~~~\\
\hline
\end{tabular}
\end{center}
\end{table}
%
It is convenient to use the terms
%
\begin{subequations}
\begin{eqnarray}
 Z_n=\sum_m S_{n,m}C_{n,m}
\end{eqnarray}
\begin{eqnarray}
 W_n(\nu)=\sum_m\frac{|S_{n,m}|^2}{\nu-\nu_{n,m}}
\end{eqnarray}
\end{subequations}
%
Then Eq.~(12) can be reduced to the form
%
\begin{equation}
 Z_N=q\,W_N\sum_{n=0}^\infty R_{N,n}Z_n 
\end{equation}
%
It is easy to show that $\,\nu_{0,0}=0,\;\,S_{0,0}=1$, that is $\,W_0=1/\nu$.
Using these features and Eq.~(14) 
one can represent the solvability condition of series (16), 
that is dispersion equation for the bunch eigentunes, 
in terms of infinite continued fraction
%
\begin{eqnarray}
 \nu-q+\frac{(q/3)^2 W_1}{1+\frac{(q/15)^2 W_1W_2}{1+\frac{(q/35)^2W_2W_3}
 {1+\dots\dots\dots}}}=0
\end{eqnarray}
%
This expression has to be truncated in reality by applying of the assumption 
$\,W_n=0\,$ at $\,n\ge n_{\rm max}$. 
Assigning corresponding truncated expression as $\,T_{n_{\rm max}}$, 
one can write the approximate dispersion equation as
%
\begin{eqnarray}
 T_{n_{\rm max}}(\nu)=0
\end{eqnarray}
%
with following recurrent relations:
%
\begin{eqnarray}
 T_n=T_{n-1}+T_{n-2}\frac{q^2W_{n-1}W_n}{(4n^2-1)^2}, \qquad (n\ge2)
\end{eqnarray}
%
and the initial conditions:
%
\begin{subequations}
\begin{eqnarray}
 T_0=\nu-q,
\end{eqnarray}
\begin{eqnarray}
\qquad T_1 = \nu-q+\left(\frac{q}{3}\right)^2\frac{3(\nu+\Delta Q)}
 {\nu(\nu+\Delta Q)-Q_s^2}
\end{eqnarray}
\end{subequations}
%
%

\section{THREE-MODE APPROXIMATION}

%

Eq.~(18) is trivial at $n_{\rm max}=0$: $\,T_0=0\,$ that is $\,\nu=q$.
It describes the wake contribution to the tune of the lowest (rigid) eigenmode 
of the bunch.
Of course, TMCI can not appear in this approximation, and the simplest equation  
to disclose it is $\,T_1(\nu)=0$ that is, in accordance with Eq.~(19)-(20),
%
\begin{eqnarray}
 (\nu-q)\left(\nu-\frac{Q_s^2}{\nu+\Delta Q}\right)=-\frac{q^2}{3}
\end{eqnarray}
%
 \begin{figure}[t!]
 \includegraphics[width=85mm]{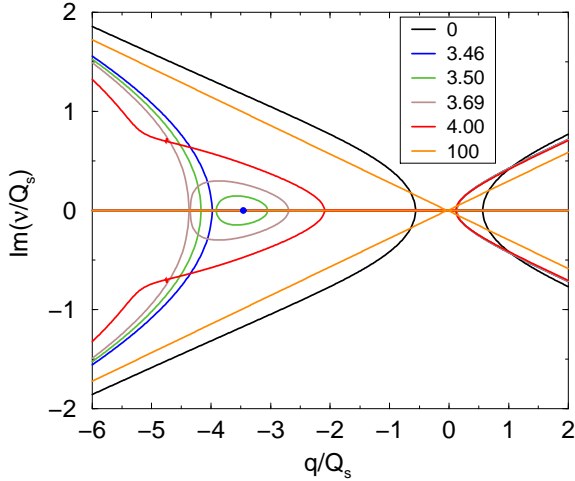}
 \caption{Imaginary part of the boxcar eigentunes
 against the wake strength at different value of the space charge tune shift. 
 Two regions of instability occur with negative wake 
 at $\,3.46<\Delta Q/Q_s<3.69\,$.}
 \end{figure}
%
This third order equation absolutely coincides with Eq.~(7.3) of 
Ref.~\cite{BA2} (without chromaticity) despite the fact that very different 
concepts have been used to derive them.
However, the examples presented in \cite{BA2} have been restricted by 
the modest space charge: $\Delta Q/Q_s<3$. 
The situation beyond this region should be explored for further development 
of the techniques.

Some solutions of Eq.~(21) are shown in Fig.~2 and Fig.~3. 
Their behavior is clear if the wake is positive: 
the TMCI threshold goes down monotonically when the space charge increases. 
The case of negative wake is more complicated and requires special comments.
%
 \begin{figure}[t!]
 \includegraphics[width=80mm]{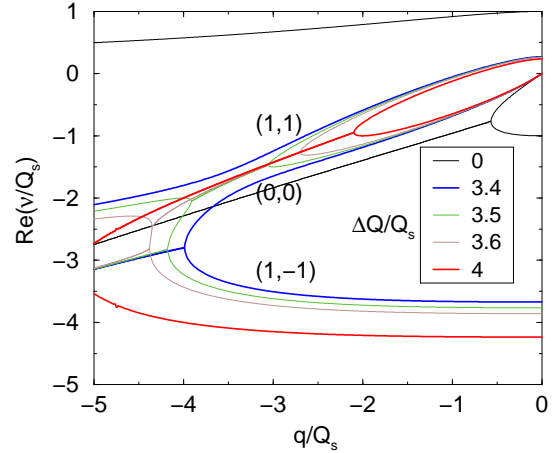}
 \caption{Real part of the boxcar eigentunes against the wake strength
 at different space charge tune shift.
 With negative wake, coalescence of the modes $\,\{0,0\}+\{1,-1\}\,$ causes 
 the instability at $\Delta/Q_s<3.46\,$ (blue lines). The switching
 on the modes $\,\{0,0\}+\{1,1\}\,$ happens at $\Delta Q/Q_s>3.69\,$(red lines).}
 \end{figure}
%

The imaginary part of the solutions is plotted against $\,q/Q_s\,$ in Fig.~2. 
According to the plot, the instability threshold is $\,|q_{\rm th}/Q_s|=0.567\,$ 
at $\,\Delta Q/Q_s=0$ (the black lines). 
Then it moves to the left (that is increases in absolute value) reaching 
$\,q_{\rm th}/Q_s\simeq -4\,$ at $\,\Delta Q/Q_s=3.46$ (blue line).
The picture crucially changes after that because new region of instability arises.
Its initial position is shown by the blue spot which quickly expands when the 
space charge tune shift increases (green and brown ovals). 
At $\,\Delta Q/Q_s=3.69$, it reaches the primary zone of instability 
which is being situated at the left of the brown parabola at this instant.
Barrier between the zones tears at higher $\,\Delta Q\,$ resulting in a single 
region of instability (red).
Its right-hand boundary goes to the right, that is the TMCI threshold goes down,
if the space charge tune shift continues to grow up.

For more explanation, real part of the same roots is plotted in Fig.~3 at $\,q<0$. 
The most important curves are shown by bold lines, and the mode index is indicated
near each of them. 
The TMCI threshold moves on the left in the beginning reaching 
$\,q_{\rm th}/Q_s\simeq -4\,$ at $\,\Delta Q/Q_s=3.46$ by a merge of the modes
\{0,0\}+\{1,-1\} (lower blue lines).  
However, the junction switches to the modes \{0,0\}+\{1,1\} at higher $\,\Delta Q\,$
resulting in the reverse motion of the threshold (higher red lines).
The intermediate positions are shown by thin dashed line.
%
 \begin{figure}[t!]
 \includegraphics[width=80mm]{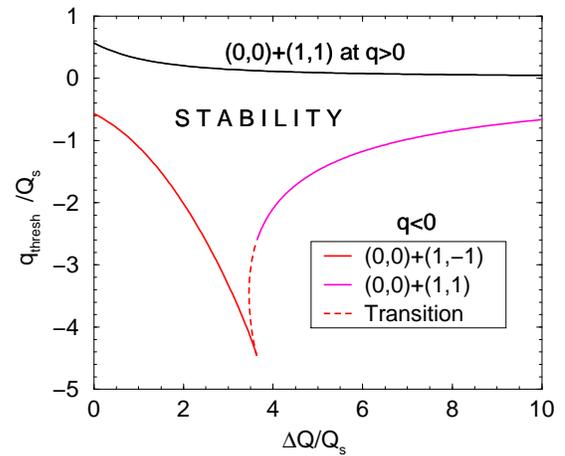}
 \caption{Stability region of the boxcar bunch at positive (black) 
 and negative (red+magenta) wakes.}  
 \end{figure}
%
 \begin{figure}[t!]
 \includegraphics[width=85mm]{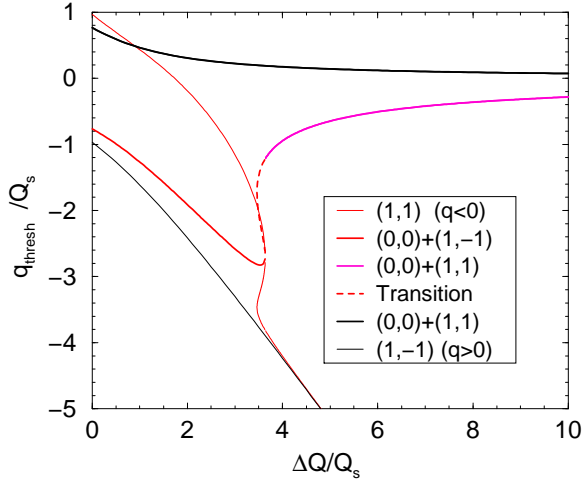}
 \caption{ The bunch eigentunes on boundaries of the stability region. 
 With positive wake, the TMCI appears as a coalescence of the modes 
 $\,\{0,0\}\,$ and  $\,\{1,1\}$ (thick black line).
 With negative wake, the modes $\,\{0,0\}+\{1,-1\}\,$ coalesce
 at $\,\Delta Q/Q_s<3.46\,$, and $\,\{0,0\}+\{1,1\}\,$ modes do it 
 at $\,\Delta Q/Q_s>3.69$.
 There is a transition region where all tunes about coincide.}  
 \end{figure}

Total region of stability in $\,(\Delta Q-q)\,$ plane is shown in Fig.~4 
with both positive and negative wakes taken into account.
The complicated shape of the red curve is explained just by the additional 
instability region which appears at $\Delta Q/Q_s=3.46\,$ and is represented
in the picture by the dashed red line.

Eigentunes on boundary of this area are shown in Fig.~5.
Generally, there are 3 real roots in the stability region but at least 2 of them 
coincides on the boundary. 
Therefore not more than 2 lines of each color are visible in the picture.
With negative wake, the modes $\,\{0,0\}+\{1,-1\}\,$ coalesce
at $\,\Delta Q/Q_s<3.46\,$, and $\,\{0,0\}+\{1,1\}\,$ modes do it 
at $\,\Delta Q/Q_s>3.69$.
There is a transition region where all tunes about coincide.
The black curves for positive wake are very simple: the top line represents 
the coalesced modes $\,\{0,0\}+\{1,1\}\,$, and the bottom line is 
tune of the last mode $\,\{1,-1\}$.    

The instability growth rate in this approximation is plotted in Fig.~6 
being shown against the wake strength at different space charge tune shifts.
The uprise of the additional instability region is seen here as well (red line). 
%
 \begin{figure}[t!]
 \begin{center}
 \includegraphics[width=80mm]{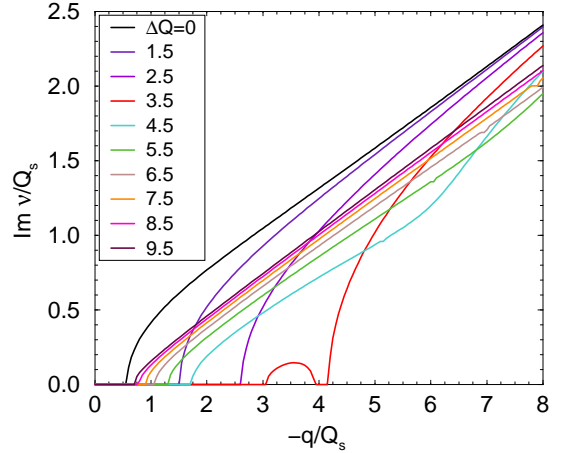}
 \end{center}
 \caption{Instability growth rate vs wake strength 
 (three-mode model, negative wake).}
 \end{figure}
%
%

\section{HIGHER APPROXIMATIONS}

%
Higher approximations should be considered to validate the three-mode model,
to establish its applicability limit, and to go beyond it.
The first step in this direction is investigation of the equation $\,T_2(\nu)=0$.
According to Eq.~(19)-(20), its expanded form  is
%
\begin{eqnarray}
 \nu-q+\frac{q^2W_1(\nu)}{9}+\frac{q^2(\nu-q)W_1(\nu)W_2(\nu)}{225}  = 0
\end{eqnarray}
%
where
%
\begin{subequations}
\begin{equation}
 W_1(\nu)=\frac{3(\nu+\Delta Q)}{\nu(\nu+\Delta Q)-Q_s^2},\qquad 
\end{equation}
\begin{equation}
 W_2(\nu)=\frac{|S_{2,-2}|^2}{\nu-\nu_{2,-2}}+\frac{|S_{2,0}|^2}{\nu-\nu_{2,0}}
+\frac{|S_{2,2}|^2}{\nu-\nu_{2,2}}
\end{equation}
\end{subequations}
%
Required parameters have to be obtained by solution of Eq.~(11) with $\,n=2$ 
which procedure is described in the Appendix.
With the notations $\,\nu_{n,m}=\hat\nu_{n,m}Q_s+\Delta Q$,
the eigentunes appear as roots of the dispersion equation
%
\begin{equation}
 \hat\nu_{2,m}(\hat\nu_{2,m}^2-4)=\frac{\Delta Q}{Q_s}\,(\hat\nu_{2,m}^2-1)
\end{equation}
%
and the formula for corresponding normalizing coefficients is
%
\begin{equation}
 |S_{2,m}|^2=\frac{5(\hat\nu_{2,m}^2-1)^2}{\hat\nu_{2,m}^4+\hat\nu_{2,m}^2+4}
\end{equation}
%
Substitution of the functions $\,W_{1-2}(\theta)\,$ into Eq.~(22) result in 
the equation
%
\begin{eqnarray}
 (\nu-q)\left(\nu-\frac{Q_s^2}{\nu+\Delta Q}\right)+\frac{q^2}{3}\nonumber
\\=-\frac{q^2(\nu-q)}{75}\left(\frac{|S_{2,-2}|^2}{\nu-\nu_{2,-2}}
 +\frac{|S_{2,0}|^2}{\nu-\nu_{2,0}}+\frac{|S_{2,+2}|^2}{\nu-\nu_{2,+2}}\right)
\end{eqnarray}
%
Its real roots can be found numerically. 
Number of the roots is 6 but only 5 different ones can appear in the boundary 
of the stability region.
It is just the way to find the threshold value of the wake
as well as corresponding eigentunes of the system with given $\Delta Q$.
The result is presented in Fig.~7 by red line. 

Similar method can be used for analysis of higher approximations as well, 
thought the corresponding formulae are more cumbersome.
Generally, it leads to the algebraic equation of power 
$\,(n_{\rm max}+1)(n_{\rm max}+2)/2\;$
where $\,n_{\rm max}\,$ is the power of the highest used
Legendre polynomial.

Some results of the calculations are represented in Fig.~7-8
where $\,q_{th}/Q_s\,$ is plotted against $\,\Delta Q/Q_s\,$ at $\,n_{\rm max}=1-10$.
It should be noted at once that the effect of positive wake almost does not depend on 
the power of the polynomial, and can be reasonably described by the three-mode Eq.~(21).
Therefore only the case of negative wake is commented below.

The curves in Fig.~7-8 are well converging, so that any approximation
$\,n_{\rm max}\ge 6\,$ can be safely used with any $\,\Delta Q\,$.
However, the lower order approximations could be sufficient if the tune shift 
is not so large.
In particular, three-mode approximation $\,(n_{\rm max}=1)\,$ provides well
accuracy at $\Delta Q/Q_s<3.5$.
The case $\,n_{\rm max}=5\,$ (21 modes) is applicable at $\,\Delta Q/Q_s<6$ 
but underestimates the TMCI threshold by $\sim$25\% at larger value of the shift. 
It is necessary to pay an attention to the narrow strip of instability 
which becomes apparent at $\,n_{\rm max}\ge 3\,$ being marked by the dashed lines 
in Fig.~7.
Its sense and cause will be explained two paragraphs below.
%
\begin{figure}[t!]
 \includegraphics[width=80mm]{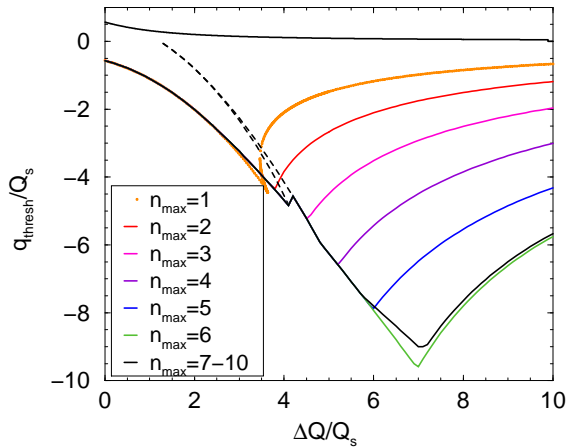}
 \caption{Stability region of the boxcar bunch at different approximations.
 Index $\,n_{\rm max}\,$ means maximal power of polynomial in the expansion. 
 The top black curve refers to positive wake with any $\,n_{\rm max}\,$, and 
 others refer to negative wakes.
 The dashed lines mark a narrow strip of instability inside the wide 
 stable area which appears at $n_{\rm max}\ge 3$.}
\end{figure}
%
\begin{figure}[t!]
 \includegraphics[width=80mm]{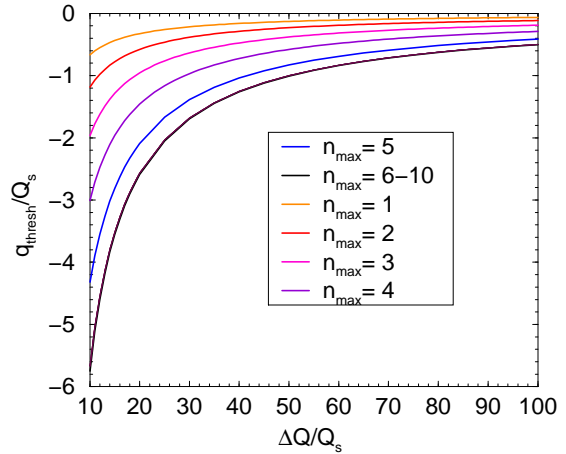}
 \caption{Same as in Fig.~7 with higher $\,\Delta Q$. The curves with 
 $\,n_{\rm max}\ge 6\,$ are indistinguishable in this plot.}
\end{figure}
%
%
 \begin{figure}[h!]
 \includegraphics[width=80mm]{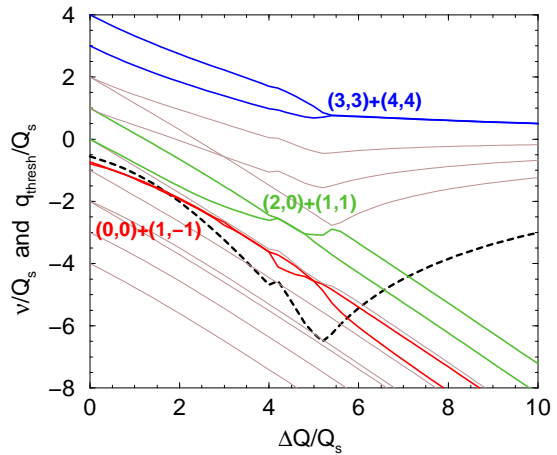}
 \caption{The bunch spectrum in the TMCI frontier at $\,n_{\rm max}=4\,$ 
 (15-mode approximation). 
 The most important modes are shown by solid bold lines and supplied with 
 inscriptions.
 The TMCI threshold is represented by dashed black line.} 
 \end{figure}
%
 \begin{figure}[h!]
 \includegraphics[width=80mm]{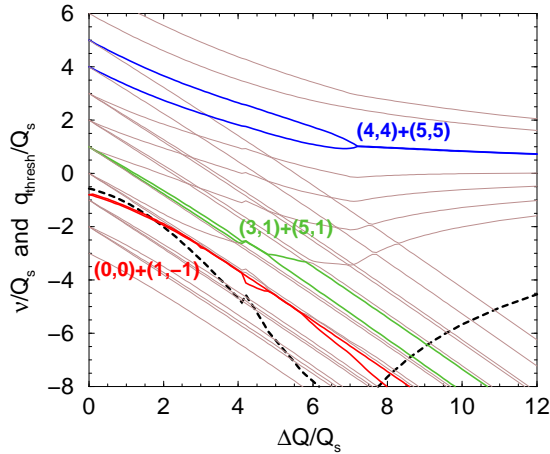}
 \caption{The same as Fig.~9 at $\,n_{\max}=7\,$ (35 modes).}
 \end{figure}
%

Important conclusions can be derived by analysis of the boundary spectrum 
that is the bunch tunes $\,\nu_{th}(\Delta Q)$ at the frontier of the TMCI zone.
Very first example has been represented in Fig.~5 by analysis of 
three-mode approximation $\,(n_{\rm max}=1)$.
It was shown that the sudden veer of the curve $\,\nu_{th}(\Delta Q)\,$ 
happens when the tune coupling switches from the modes $\,\{0,0\}+\{1,-1\}\,$ 
to $\,\{0,0\}+\{1,1\}\,$

The case $\,n_{\rm max}=4\,$ is represented in Fig.~9.
All 15 tunes of this approximation are shown in the graph, and 
the most important of them are accentuated by the bold lines and inscriptions.
In the beginning, the TMCI occurs due to coupling of the modes   
$\,\{0,0\}$ and $\{1,-1\}\,$ which are shown by red  lines.
Then it passes through the $\,\{2,0\}+\{1,1\}\,$ coalescence (green lines) 
to the final combination $\,\{3,3\}+\{4,4\}\,$ (blue).
Each step is followed by a veer of the threshold curve (dashed black line). 
Other short-time conjunctions or crossings of the spectral lines are seen in 
the picture as well.
They can be responsible for petty areas of instability one of which 
is marked in Fig.~7 by dashed lines. 
Perhaps, there are other similar zones but they are elusive because of small size. 

Similar behavior of the spectrum is demonstrated in Fig.~10 
where the case $\,n_{\rm max}=7\,$ is considered.
However, there is an important difference from previous example 
because the ultimate coupling of the modes $\,\{4,4\}+\{5,5\}\,$ 
happens inside the spectrum, and there are 2 tunes above the junction. 
It confirms that the saturation is achieved at $\,n_{\rm max}=6$, 
and additional basis vectors no longer affect the mode coupling.

Summarizing this section, one can propose following approximate formulae 
for negative wake:
%
\begin{equation}
 q_{th} \simeq \bigg\{ \begin{array} {ccc}
   -\sqrt{(0.57Q_s)^2+(1.3\Delta Q)^2}&\quad{\rm at}\quad&\;\Delta Q/Q_s<7\\
   -50 Q_s^2/\Delta Q                 &\quad{\rm at}\quad&\;\Delta Q/Q_s>7
\end{array}
\end{equation}
%
\\
The case of positive wake is much simpler: TMCI is caused by a coupling 
of the modes $\,\{0,0\}+\{1,1\}\,$ at any $\,\Delta Q\,$, 
and three-mode approximation is sufficient for the calculation resulting in the 
threshold formula
%
\begin{equation}
 q_{th}\simeq 0.57\left(\sqrt{Q_s^2+\frac{\Delta Q^2}{4}}
 -\frac{\Delta Q}{2}\right)
\end{equation}
%
Eq.~(27)-(28) are the fits of Fig.~7-8 providing the accuracy not worse of 
15\% at any tune shift.
%

\section{Discussion}

%
Comparison of the results with published papers is the subject of this section.

Fig.~1 of Ref.~\cite{BL1} is reproduced in upper Fig.~11 of this paper.
The TMCI growth rate is plotted in the graph against the tune shift at fixed 
wake strength which corresponds to $\,q=-1.13\,$ in my terms.
The same function is plotted in lower Fig.~11 which is calculated by solution 
of Eq.~(21). 
It would be possible to tell about amazing similarity of the pictures 
if not the difference of the horizontal scales. 
According to the upper graph, the space between two instability regions is
$\,\Delta Q/Q_s=1.4\,$ whereas it should be about 3 times more by the bottom plot.	
Another important thing is that the second region of instability 
appears only in the high-mode approximation $\,m_{\rm max}=10$ in the upper
picture.
However, it would be situated far to the right $\,(\Delta Q/Q_s>\sim 45)\,$ 
if corresponding high-order approximation is used for the lower graph.
Therefore, one can declare only qualitative agreement with this result 
of the paper \cite{BL1}.
Another model of Ref.~\cite{BL1} is a hollow bunch in the square potential well.
In contrast with previous, it predicts a monotonic growth of the TMCI threshold at 
increasing tune shift. 

There is a better agreement with Ref.~\cite{BL2} where results of numerical 
simulations are represented.
Shape of the threshold curve with  negative wake is very similar to Fig.~7
at $\,\Delta Q/Q_s\le 6$, and there is a rather good quantitative agreement.
For example, both papers give $\,q_{th}/Q_s\simeq -6.5$ at $\,\Delta Q/Q_s=5$.
Thresholds of positive wakes are in a full consent as well.
%
 \begin{figure}[t!]
 \includegraphics[width=80mm]{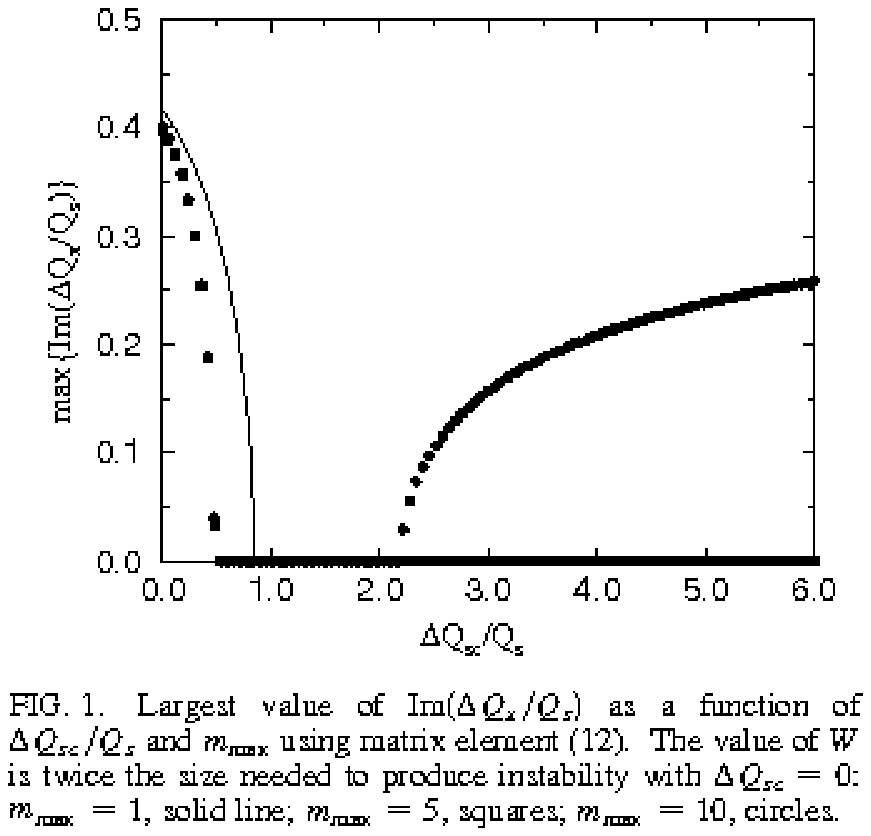}
{Copy of Fig.~1 \cite{BL1}}
 \end{figure}
%
 \begin{figure}[t!]
 \includegraphics[width=80mm]{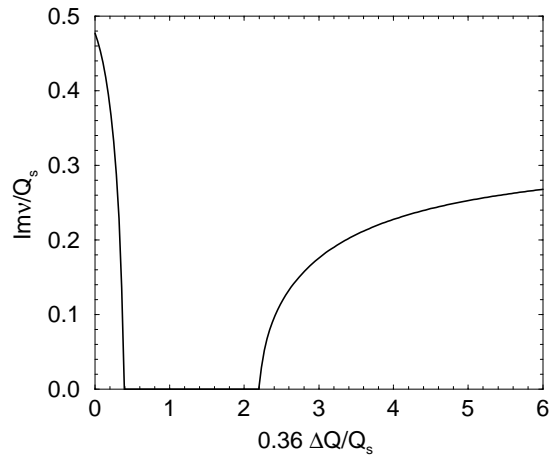}
 \caption{TMCI growth rate against space charge tune shift by \cite{BL1} (top) 
 and by Eq.~(21) (bottom), at fixed wake strength $q=-1.13$.}
 \end{figure}
%
%
 \begin{figure}[b!]
 \includegraphics[width=80mm]{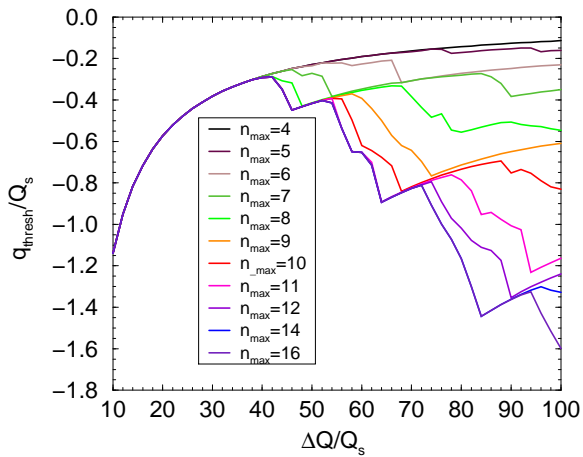}
 \caption{TMCI threshold with the limiting parameters described by Eq.~(29).
  Top black line: $\,n_{\rm max}=2-4$, further $\,n_{\rm max}=5-16,\;\Delta n=1$.}
 \end{figure}
%

The case $\,\Delta Q\rightarrow\infty\,$ is more complicated and requires 
a careful consideration. 
It follows from Fig.~7-8 and Eq.~(27)-(28) that the TMCI threshold tends to 0 
in this limit.
In contrast, it was argued in papers ~\cite{BUR} and \cite{BA1} that negative 
wake can't cause TMCI at all at such conditions ("vanished TMCI \cite{BUR}").
It was noted in the Introduction that this statement is questionable 
because negative multipoles have been actually excluded from the consideration 
in mentioned papers.
However, we can see now that this objection is certainly essential only 
at $\,\Delta Q/Q_s<\sim 7\,$ and maybe is not so important at higher $\,\Delta Q\,$
where the instability is caused just by coupling of positive multipoles 
(see Fig.~10).

It should be noted in this connection that central equations of the quoted papers 
could be solved for the boxcar bunch by the method which is developed 
in this paper.
It would result in Eq.~(17) with additional conditions:
%
\begin{subequations}
\begin{equation}
 \nu_{n,n} = \frac{Q_s^2 n(n+1)}{2\Delta Q}		  
\end{equation}
\begin{equation}
 |S_{n,m}|^2 = (2n+1)\,\delta_{n,m},\qquad 
\end{equation}
\end{subequations}
%
which are simply the asymptotic values of these parameters at 
$\,\Delta Q\rightarrow\infty$.
However, truncation of significant part of the spectrum produced by Eq.~(29b)
crucially changes the result.
It is illustrated by Fig.~12 which is obtained by solution of Eq.~(18) with 
assumptions (29) at different $\,n_{\rm max}$.   
The graph drastically differs from Fig.~8 though both of them are calculated by the 
same method but with/without the truncation.
Despite the poor convergence of the curves in Fig.~12, trend to the threshold 
growth is evident. 
However, it is seen now that this conclusion is unreliable and should be rejected
if tune shift exceeds synchrotron tune on one or two orders of value.   
It confirms the supposition which can be expressed first by Burov \cite{BUR}.

%

\section{Conclusion}

%
Being stable in themselves, eigenmodes of the boxcar bunch with space charge
form convenient and effective basis for investigation of the bunch instability 
with space charge and wake field.	 
Dispersion equation derived by this method is represented in the form of 
infinite continued fraction as well as in the form of recursive relation with 
arbitrary number of the basis vectors included.

It is shown that effect of the space charge on the instability essentially depends 
on the wake sign.
TMCI threshold of positive wake goes down monotonously when the space charge 
tune shift increases as it is approximately described by Eq.~(28).
The negative wake  threshold goes up in absolute value if $\,\Delta Q/Q_s<7\,$, 
and goes down at higher $\,\Delta Q/Q_s$, Eq.~(27).
Such a complicated behavior occurs because the coalescence of different modes 
is responsible for the instability at different $\,\Delta Q/Q_s$.
If the wake is negative, the multipoles $\,m=0\,$ and $\,m=-1\,$ are coupled at 
lower $\,\Delta Q\,$ and $\,m=6-7$ at higher one. The rearrangement comes with an 
abrupt dog-leg of the threshold curve.
With positive wake, the multipoles $\,m=0\,$ and $\,m=1\,$ are responsible for the 
instability at any tune shift resulting in monotonous decrease of the 
threshold. 
Additional investigation of variable wakes looks to be important.  


\section{Acknowledgments}

FNAL is operated by Fermi Research Alliance, LLC under contract 
No. DE-AC02-07CH11395 with the United States Department of Energy.


\section{Appendix}

%
Using the notation
$$
 \hat\nu_{n,m}=\frac{\nu_{n,m}+\Delta Q}{Q_s},\quad
 \hat\Delta Q =\frac{\Delta Q}{Q_s},\quad
 P_n(\theta) = \sum_{l=0}^n p_{nl}\theta^l
 \eqno(A1)
$$
one can write Eq.~(11) in the form
$$
 \hat\nu_{n,m}Y_{n,m}+i\,\frac{\partial Y_{n,m}}{\partial\phi}=
 S_{n,m}\Delta\hat Q \sum_{l=0}^np_{n,l}(A\cos\phi)^l
 \eqno(A2)
$$
Its solution is
$$
 Y_{n,m}=S_{n,m}\Delta\hat Q\sum_{k=-n}^n\frac{\exp{ik\phi}}{\hat\nu_{n,m}-k}
 \sum_{j=0}^n U_{n,k,j}A^{k+2j} 
 \eqno(A3)
$$
where
$$
 U_{n,k,j}=\frac{p_{n,k+2j}}{2^{k+2j}}{k+2j\choose j} \times
 \bigg\{\begin{array}{ll} 1 {\rm\;\;at}\;\;k+j\ge 0     \\
                          0 {\rm\;\;at}\;\;k+j <  0     \end{array}
 \eqno(A4)
$$
This function should satisfy normalization condition represented by Eq.~(6) with 
$\,j\equiv\{n,m\}\,$ and distribution function (10a).
The substitution results in the relation: 
$$
 1=\sum_{k=-n}^n\frac{S_{n,m}^2\Delta\hat Q^2}{(\hat\nu_{n,m}-k)^2}
 \sum_{j_1=0}^n\sum_{j_2=0}^n U_{n,k,j_1} U_{n,k,j_2}
 \overline{A^{2(k+j_1+j_2)}}
 \eqno(A5)
$$
where $\,\overline {A^{2j}}\,$ is the amplitude power averaged over the 
distribution:
$$
 \overline{A^{2j}}=\int_0^1\frac{A^{2j+1}\,dA}{\sqrt{1-A^2}}=
 \sum_{l=0}^j {j\choose l} \frac{(-1)^l}{2l+1}
 \eqno(A6)
$$
In principle, involved eigentunes $\,\nu_{n,m}\,$ could be obtained by 
substitution of Eq.~(A3) into Eq.~(3) with the functions $\,Y\,$ and $\,\bar Y\,$
being taken from this Appendix.
Because similar calculation has been actually accomplished in Ref.~\cite{SAH}, we
represent here only the resulting equation for the eigentunes:
\\
\\
Lower powers
$$
 \hat\nu_{0,0}=\Delta\hat Q,\qquad 
 \hat\nu_{1,\pm1}^2-1=\Delta\hat Q\hat\nu_{1,\pm1}
 \eqno(A7)
$$
Higher even powers
$$
 \hat\nu_{n,m}[\hat\nu_{n,m}^2-4]\dots[\hat\nu_{n,m}^2-n^2]
$$
$$
=\Delta\hat Q [\hat\nu_{n,m}^2-1]\dots[\hat\nu_{n,m}^2-(n-1)^2]
\eqno(A8)
$$
Higher odd powers
$$
 [\hat\nu_{n,m}^2-1]\dots[\hat\nu_{n,m}^2-n^2]
$$
$$
=\Delta\hat Q\hat\nu_{n,m} [\hat\nu_{n,m}^2-4]\dots[\hat\nu_{n,m}^2-(n-1)^2]
\eqno(A9)
$$
\\
Some of the values $\,\nu_{n,m}/Q_s=\hat\nu_{n,m}-\Delta\hat Q\,$ are plotted 
in Fig.~1.
Factors $S_{n,m}^2$ can be found from Eq.~(A5) with the known eigentunes 
substituted. 
Some results are represented below:
$$
 S_{0,0}^2=1,\qquad S_{1,\pm1}^2=\frac{3\hat\nu_{1,\pm1}^2}{\hat\nu_{1,\pm1}^2+1}\,,
\eqno(A10)
$$ 
$$
 S_{3,m}^2=\frac{5(\hat\nu_{2,m}-1)^2}
 {\hat\nu_{3,m}^4+\hat\nu_{2,m}^2+4}\,,
\eqno(A11) 
$$
$$
 S_{3,m}^2=\frac{7\hat\nu_{3,\pm1}^2(\hat\nu_{3,m}-4)^2}
 {\hat\nu_{3,m}^6-2\hat\nu_{3,m}^4+13\hat\nu_{2,m}^2+36}\,,
 \eqno(A12) 
$$
etc.

%


\end{document}